\documentclass[11pt]{article}
\usepackage[margin=1in]{geometry}
\usepackage{graphicx}

\def\beq{\begin{equation}}
\def\eeq#1{\label{#1}\end{equation}}
\def\eeqn{\end{equation}}

\def\beqa{\begin{eqnarray}}
\def\eeqa#1{\label{#1}\end{eqnarray}}
\def\eeqan{\end{eqnarray}}

\let\bar=\overbar

\def\Dslash{\not{\hbox{\kern-4pt $D$}}}
\def\dslash{\not{\hbox{\kern-2pt $\del$}}}

\def\msb{{\bar{\ssstyle M \kern -1pt S}}}

\def\Title#1{\begin{center} {\Large {\bf #1} } \end{center}}
\def\Author#1{\begin{center} {\normalsize {\sc #1} } \end{center}}
\def\Institution#1{\begin{center} {\normalsize {\it #1} } \end{center}}
\def\Abstract#1{\noindent {\normalsize {\bf Abstract:} {\normalfont #1}}}
\def\Conference{\vspace{4mm}\begin{raggedright} {\normalsize {\it Talk presented at the 2019 Meeting of the Division of Particles and Fields of the American Physical Society (DPF2019), July 29--August 2, 2019, Northeastern University, Boston, C1907293.} } \end{raggedright}\vspace{4mm}}

\begin{document}

%%%%%%%%%%%%%%%%%%%%%%%%%%%%%%%%%%%%%%%%%%%%%%%%%%%%%%%%%%%%%%%%%%%%%%%%%%%
%
% TITLE, AUTHOR, INSTITUTION, ABSTRACT ==> UPDATE
% 
%%%%%%%%%%%%%%%%%%%%%%%%%%%%%%%%%%%%%%%%%%%%%%%%%%%%%%%%%%%%%%%%%%%%%%%%%%%

\Title{Liquid Argon TPC Trigger Development with SBND}

\Author{Georgia Karagiorgi, for the SBND Collaboration}

\Institution{Department of Physics\\ Columbia University, New York, NY 10027, USA}

\Abstract{The Short Baseline Near Detector (SBND) is a 112 ton active mass liquid argon time projection chamber (LArTPC) that will begin operations in the Booster Neutrino Beamline at Fermilab in 2020. Its main physics goals include high-statistics measurements of neutrino-argon interaction cross-sections and searches for sterile neutrino oscillations as part of three LArTPCs that make up the Short Baseline Neutrino (SBN) Program at Fermilab. In addition, SBND serves as an R\&D platform for future LArTPC detectors such as those employed by the Deep Underground Neutrino Experiment (DUNE). One of the technical challenges of DUNE that SBND aims to address is that of efficient self-triggering utilizing TPC signal information. Such capability will enable searches for rare processes in the DUNE far detector, for example neutrino interactions from a potential galactic supernova burst, or proton decay. These proceedings describe the SBND TPC readout system and ongoing R\&D efforts to develop and demonstrate efficient TPC-based self-triggering.}

\Conference

%%%%%%%%%%%%%%%%%%%%%%%%%%%%%%%%%%%%%%%%%%%%%%%%%%%%%%%%%%%%%%%%%%%%%%%%%%%
%
% MAIN TEXT ==> UPDATE
% 
%%%%%%%%%%%%%%%%%%%%%%%%%%%%%%%%%%%%%%%%%%%%%%%%%%%%%%%%%%%%%%%%%%%%%%%%%%%

\section{Introduction}

The use of liquid argon time projection chamber (LArTPC) detectors for neutrino physics has grown rapidly over the last two decades, culminating into the proposed Deep Underground Neutrino Experiment (DUNE) \cite{duneidr,dunetdr}. DUNE will use four LArTPCs, each 10~kilotons in fiducial mass, in order to study three-neutrino oscillations, with the ultimate goals of discovering charge-parity violation in the lepton sector and determining the neutrino mass hierarchy. The large mass and deep underground location of the four DUNE far detector LArTPCs also render the experiment sensitive to potential new physics manifesting as rare events, such as proton decay, neutron-antineutron oscillation, and to low-energy neutrino interactions in case of a once-in-a-century galactic supernova burst. 

Accessing this rare physics processes will require continuous processing of the full data rate of the DUNE far detector, which is computationally challenging. To set the scale, just one of the four DUNE far detector LArTPCs will be streaming out 1.15~terabytes of data per second.\footnote{This rate assumes a single-phase LArTPC with 150 anode plane arrays, each with 2,560 wires independently read out and digitized at 2~MHz with a 12-bit ADC. Similar rates are expected for a dual-phase LArTPC.} In preparation for DUNE, LArTPC readout and trigger development is ongoing with currently-operating smaller-scale (20-500 times smaller than DUNE) LArTPCs, including MicroBooNE \cite{ubdet} and ProtoDUNE-SP \cite{protodune}. These proceedings summarize related trigger development efforts with the upcoming Short Baseline Near Detector (SBND) \cite{Antonello:2015lea}.

\section{LArTPC Triggering}

LArTPCs such as those employed by DUNE, MicroBooNE, ProtoDUNE-SP, and SBND, alike, work by stereoscopically imaging particle interactions taking place randomly over a large detector volume. The imaging is achieved by drifting ionization charge that is liberated by charged particles traversing a contiguous liquid argon volume toward a set of sensor wire arrays. The drifting of ionization electrons takes place over a relatively long distance (or, equivalently, over a timescale of order milliseconds) under the influence of a constant, high electric field. The drifted ionization electrons lead to induced or collected charge across the highly-segmented arrays of sensor wires; the induced or collected charge signal on each wire is then shaped, amplified, and sampled (synchronously for all wires) at, typically, 2~MHz. This results effectively into the continuous streaming of two-dimensional (channel versus time) projections of activity within the physical LArTPC active volume. The continuous digitized waveforms (one per wire channel) are then propagated downstream to a data acquisition system (DAQ) for further processing in real time or online, and a subset of the waveform data is recorded offline for further analysis.

Because of the relatively large drift timescale (of order milliseconds) and consequently the long integration times involved in reading out information for any given interaction within the detector volume, a data-driven data selection (trigger) decision requires:
\begin{itemize}
    \item large buffering to temporarily hold a copy of the data while the decision is being made; and
    \item fast and efficient data processing so that the trigger decision is completed within a time interval granted by the buffering of the data copy.
\end{itemize}
While for most physics signal topologies a sufficiently long buffering timescale corresponds to O(10) milliseconds, supernova burst signatures are extended in time, lasting multiple seconds. As such, orders of magnitude more buffering and processing may be needed for a supernova burst trigger, as it may require integration of information collected over up to 10 seconds in order to make an efficient data-driven supernova burst trigger decision. The need to process O(10) seconds worth of data for such trigger decision weighs in both background activity in the detector and the physics details of a supernova burst, and in particular it exploits the large multiplicity of neutrino interactions expected over that period of time in case of a galactic supernova. For reference, about 50 neutrino interactions are expected in 10~kilotons of liquid argon, most arriving within a few seconds of each other, for a supernova burst at 25~kpc. This distance roughly corresponds to the far edge of our galaxy. 

In the case of DUNE, a supernova burst trigger benefits from the sheer size of the detector as well as the shielding provided against cosmogenic backgrounds by a mile of overburden rock. The most dominant background to low-energy (a few to a few tens of MeV) neutrino interactions in DUNE is expected to be from pileup of intrinsic radioactivity within the detector itself and electronics noise. Efforts are ongoing within DUNE \cite{dunetdr} in order to develop trigger strategies for supernova burst detection, and using ProtoDUNE-SP as a development and testing platform. 

In the case of SBND, the small size of the detector (112 tons of active mass) implies two orders of magnitude fewer neutrino interactions present for any given supernova burst, as compared to DUNE. Furthermore, SBND's on-surface location, and in particular its minimal shielding and exposure to a high rate of cosmic ray muons and other cosmogenic activity, make the possibility of self-triggering on a galactic supernova burst with SBND a very challenging task, at best. Nevertheless, SBND provides another opportunity for trigger development, benefiting from the successfully demonstrated TPC readout design it will employ, which is based on that of the already-operating MicroBooNE experiment \cite{ubdet}. The SBND TPC readout system is described in the following section. 

\section{The SBND TPC Readout System}

\begin{figure}[htb]
\centering
\includegraphics[width=6in, trim=50 50 50 50]{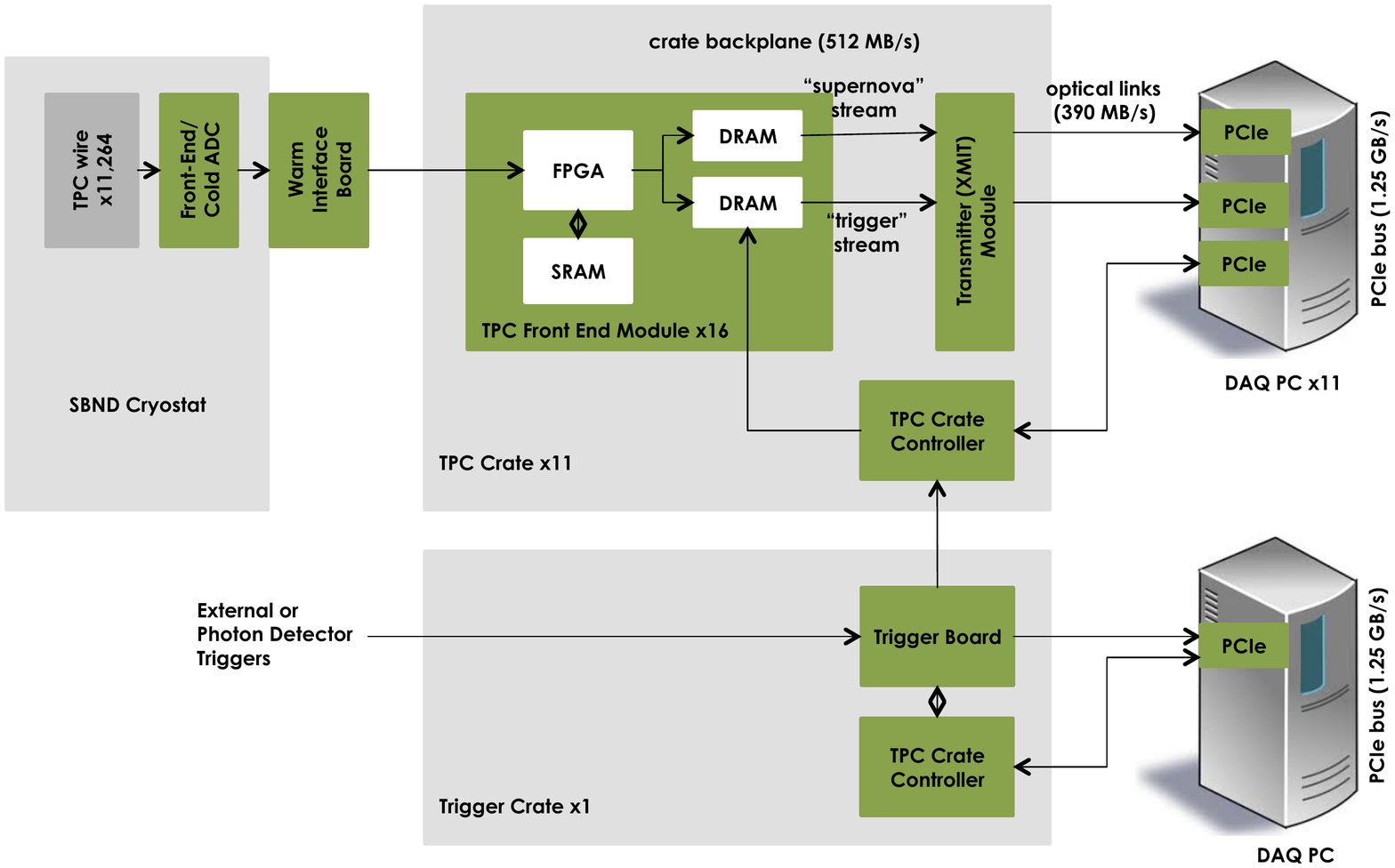}
\caption{The SBND TPC readout system, represented by the components in the two grey blocks in the center of the figure. To the left, the system interfaces with the TPC Cold Electronics System of SBND; to the right, it interfaces with the back-end part of the SBND DAQ system.}
\label{fig:readout}
\end{figure}

The SBND TPC Readout System is responsible for reading out and processing signals from SBND's 11,264 TPC wires. The wires are subdivided into four anode plane arrays (APAs) in the detector; each APA holds 2,816 wires distributed among two induction wire planes and one collection wire plane. The four APAs view four independent detector regions, each with the same drift dimension that corresponds to a 1.28~ms drift. The wire signals are digitized independently, on a per wire channel basis, using the Cold Electronics System described in \cite{shanshandpf}; digitization is carried out inside the detector, in liquid argon, in order to minimize noise, and at 2~MHz, with a 12-bit ADC. Digitized wire data from the cold electronics system are sent via optical fibers to the TPC readout system, which is a custom back-end electronics system designed by Nevis Laboratories, Columbia University. The TPC readout system and TPC data flow are illustrated in Fig.~\ref{fig:readout}. The system is distributed over eleven crates, each housing a crate controller, a transmitter (XMIT) module, a timing distribution card (not shown in the figure), and 16 Front End Modules (FEMs). A trigger board, housed in one of the eleven crates, issues or receives and distributes external trigger signals to the entire TPC readout system.

%%%%%%%%%%%%%%%%%%%%%%%%%%%%%%%%%%%%%%%%%%%%%%%%%%%%%%%%%%%%%%%%%%%%%%%%%
%%
%%   use this format to include an eps/pdf figure into your paper
%%
%%%%%%%%%%%%%%%%%%%%%%%%%%%%%%%%%%%%%%%%%%%%%%%%%%%%%%%%%%%%%%%%%%%%%%%%%%%

The FEMs are responsible for the bulk of the digital data processing. Each FEM receives digitized data from the detector cryostat for a group of 64 TPC wire channels, with data arranged in channel and then time order. The FEM first re-arranges the data it receives in time and then channel order and into synchronous (across the TPC readout) intervals of 1.28~ms, referred to as ``frames.'' The system then duplicates the data into two streams, and further processes and reads out each data stream in parallel and independently of each other. The first parallel stream is referred to as the ``trigger stream.'' It is an externally-triggered data stream in which finite-length TPC waveform data is read out upon receipt of an external trigger, prior to its distribution to the SBND back-end DAQ. An example such trigger is a neutrino beam trigger from the Fermilab accelerator. Upon receipt of any external trigger, 3.84~ms' worth of continuous waveform data for all channels, starting 1.28~ms before the absolute trigger time, gets collected and losslessly compressed in each FEM using Huffman encoding, and then sent to the back-end DAQ to be built as part of a triggered event.

\begin{figure}[htb]
\centering
\includegraphics[width=4.8in, trim=50 120 0 100]{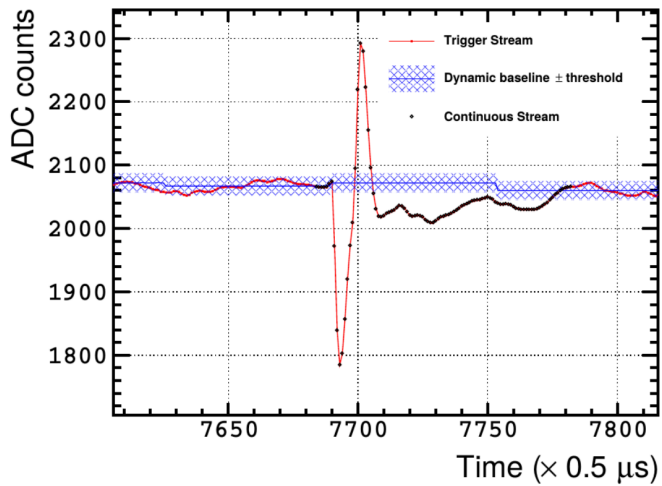}
\caption{ROI-finding scheme for SBND's continuous ``supernoval'' readout stream. In this example, shown for an induction channel waveform, after dynamic baseline estimation, represented by the blue solid line, a threshold is applied above and below baseline, and threshold crossings are identified, representing an ROI. In this waveform, the threshold is met between samples 7690 and 7705, and those samples along with a (configurable) number of pre- and post-samples relative to those are also recorded as part of the ROI data, represented by the waveform region in black. A coincident trigger would result in the }
\label{fig:ROI}
\end{figure}

The second parallel stream is referred to as the ``supernova stream.'' It is a continuous, trigger-less data stream, in which TPC data is continually rearranged into frames, with lossy data reduction applied to it prior to its distribution to the SBND back-end DAQ for semi-permanent storage. The supernova stream processing works by applying zero-suppression on a per-channel basis. The zero-suppression scheme is identical to that already employed by MicroBooNE, described in \cite{ubpubnote}, and illustrated in Fig.~\ref{fig:ROI}. Regions-Of-Interest (ROIs) are extracted from each wire's continuous waveform data by applying a configurable, channel-dependent threshold above and/or below channel baseline. The channel baseline can be configured as a static variable throughout each run (this is presently the chosen MicroBooNE configuration), or computed dynamically, by tracking baseline and baseline variance changes over successive time windows on a per channel basis and in real time. The ROIs are further compressed using lossless (Huffman) compression. The combination of zero-suppression and Huffman compression is targeting an overall data reduction factor of 80, while preserving signals associated with ionization charge deposits of minimum ionizing and other particles. ROIs are finally distributed to back-end DAQ, and saved semi-permanently in offline disk (for up to 2 days). Upon an external SuperNova Early Watch System (SNEWS) alert \cite{Antonioli:2004zb}, the semi-permanently saved ROIs can be retrieved, permanently saved, and analyzed offline to find any potential supernova neutrino interactions.

%%%%%%%%%%%%%%%%%%%%%%%%%%%%%%%%%%%%%%%%%%%%%%%%%%%%%%%%%%%%%%%%%%%%%%%%%
%%
%%   use this format to include an eps/pdf figure into your paper
%%
%%%%%%%%%%%%%%%%%%%%%%%%%%%%%%%%%%%%%%%%%%%%%%%%%%%%%%%%%%%%%%%%%%%%%%%%%%%

The trigger stream can accommodate a trigger rate of up to 20~Hz, and guarantees minimum-bias data for all beam-related and external-trigger physics measurements of SBND. The ``supernova stream'' can handle input data rates of approximately 50~gigabytes per second, and, given a compression factor of 80, it guarantees the recording of hundreds of seconds of supernova-burst-coincident data upon receipt of an external SNEWS alert. 

Because the supernova stream is completely decoupled and runs independently of the trigger stream, it also provides a useful platform for studying on- or off-beam signals, including low-energy signals, and in particular for understanding the impact of data reduction and data selection schemes on associated low-energy and other physics.

\section{SBND Trigger Development Toward DUNE}

Following the MicroBooNE supernova stream development efforts as an example, SBND can also be used to benchmark the performance of TPC data reduction and data selection algorithms on low-energy activity. For example, MicroBooNE has recently demonstrated that zero-suppressed collection plane data can be used to obtain comparable reconstruction performance for low-energy electrons from cosmic ray muon decay (Michel electrons) as with non-zero-suppressed data \cite{ubpubnote}. This builds confidence on the applicability of similar data processing techniques for supernova neutrino searches, where the characteristic signal is an electron of similar energy as a Michel electron.

More relevant to the DUNE DAQ and in particular the DUNE data selection design, SBND's continuous readout stream can be used to extract trigger primitive information which can be considered as input for a fast (online or real-time) trigger decision. Trigger primitives for DUNE are defined as summaries of per-channel waveform data. They can be implemented, for example, as summaries of threshold-based hit information including: hit channel number, time of threshold, time over threshold, and amplitude or integrated ADC charge. The DUNE data selection design presently considers only collection plane trigger primitives. Those form the basis of a higher level trigger decision; more specifically, the aggregation of certain trigger primitives within a detector sub-component (e.g.~an APA) is used to form ``trigger candidates'', and subsequent correlation of those within an APA or an entire detector module is used to form a trigger decision, prompting the readout of the detector module \cite{dunetdr}.

SBND offers a valuable platform for testing and further developing the DUNE data selection design. ROI finding in SBND's continuous readout stream will be used as a basis for trigger primitive generation, and subsequently higher-level (trigger candidate and trigger decision) data selection algorithms will be implemented and used to exercise and test a full data selection chain. The SBND TPC readout system provides flexibility for doing so either in software or in hardware (in the XMIT module firmware). Preliminary studies carried out using offline analysis of MicroBooNE data show that trigger primitive information extracted from collection plane ROIs such as those expected in SBND is promising as the basis for a TPC activity- and topology-based trigger, for example a Michel trigger \cite{nsfreu}. Such trigger algorithms are currently being developed for online implementation in ProtoDUNE-SP this fall, and for real-time implementation in the transmitter module firmware of SBND in summer 2021. 

SBND's readout design offers the possibility of exploring trigger primitive generation and utilization for a higher-level data selection using not only collection plane but also collection and/or induction plane information, interchangeably. Furthermore, unlike MicroBooNE, where the mapping of the physical TPC wires to FEM boards is relatively inflexible, the optical interface between the cold ADC electronics and warm digital readout electronics in SBND provides flexibility for regrouping inputs to FEMs to correspond to physically contiguous sets of channels, so that spatial and time correlations can be better exploited at the hardware level for a data selection decision. This will take place at the transmitter module level, which collects information in groups of up to 1,024 independent channels.

The availability of a flexibly modifiable real-time ROI stream at SBND will also serve the development of other novel LArTPC trigger techniques, e.g.~ones involving online or real-time image processing. E.g., the use of Deep Neural Networks (DNNs) has been proposed for LArTPC image classification for triggering purposes \cite{nysds}. This can be done online in GPUs or CPUs, or using hardware acceleration in field programmable gate arrays (FPGAs) for real-time classification. Each implementation option comes with advantages and disadvantages in terms of power and resource utilization and speed of inference. SBND can be used to demonstrate and compare different implementations in situ with real LArTPC data.

\section{Summary and Conclusions}

The SBND detector is currently under construction, with most major detector components already delivered to Fermilab and under installation at the SBND building. The SBND TPC construction is on schedule to be completed in fall 2019, along with TPC readout electronics delivery and installation at Fermilab. TPC readout electronics commissioning is expected to be completed by summer 2020, in time for SBND detector commissioning and first neutrino data in 2021.

The SBND TPC readout system will employ a dual readout stream, which guarantees lossless readout of beam and other externally-triggered events for SBND's physics program, as well as continuous detector readout enabling offline searches for supernova neutrinos. The common readout design aspects between the SBND continuous readout and the future DUNE DAQ offer a unique opportunity for LArTPC trigger development that will have direct relevance to DUNE. Efforts to implement real-time triggering in the SBND TPC readout, following the DUNE data selection prescription, are currently ongoing, targeting deployment and testing in situ at SBND in summer 2021, during the first scheduled accelerator beam shutdown period for the experiment.

\section*{Acknowledgements}
These proceedings represent the work of SBND collaborators, many of whom are also collaborators on MicroBooNE and DUNE. Much of the groundwork for the TPC trigger development effort on SBND is also taking place within the MicroBooNE and DUNE collaborations. Particular thanks are due to J.~Crespo-Anadon for contributions to these development efforts, and for Fig.~\ref{fig:ROI} in these proceedings. This work is supported by the National Science Foundation under Grant No.~PHY-1753228.

\end{document}